\documentclass[12pt,twoside]{article}

\usepackage{fleqn,espcrc1}
\usepackage{graphicx}

\title{Optical potentials for the antiproton nucleus interactions}

\author{S. Wycech \address{So{\l}tan 
Institute for Nuclear Studies, PL-00-681 
Warsaw, Poland}\thanks{Representing PS209}}

\begin{document}
\maketitle

\pagestyle{empty}
     
\begin{abstract}
The nuclear interactions of atomic and low energy antiprotons are studied. 
Measurements of level shifts and widths in the lightest elements are 
analyzed and compared with new results obtained in heavy nuclei. Simple 
geometric properties of  $\bar{p}$ nucleus interactions are demonstrated. 
Upon this background one finds some anomalies that indicate strong energy 
dependence in the subthreshold $\bar{p}$ nucleon interactions. The use of 
of  $\bar{p}$ in studies of  the nuclear surface is briefly discussed. 
\end{abstract}

\section{INTRODUCTION} 
 
The study of  antiprotonic atoms and low energy scattering serves a 
triple purpose:

(a) {\sl to check some properties of} $\bar{p}$  {\sl nucleon interactions}

(b) {\sl to learn about the structure of the nuclear surface}

(c) {\sl to find some exotic phenomena as $\bar{p}$ - nucleus quasi-bound states}.

(a). In practice this possibility is limited  to two simple 
systems:  nucleon and deuterium, nevertheless we argue below that
heavier atoms also offer some advantages if there exists a subthreshold 
$\bar{N}N$ resonance. 

(b). Several methods have been used for this purpose, each
of them gives  information in regions roughly 1-4 fm beyond 
the half-density radius. One method studies the X-ray cascade and extracts
the atomic level shifts and widths. Until now, the information obtained
in this way has been  limited to at most one level shift $\Delta E$ 
and two level widths  $\Gamma$ for a given atom \cite{ROB77}.
The recent CERN  experiment \cite{TRZ00} 
detects more transitions,  improves the precision, resolves the 
fine structure and allows for several shifts and widths per atom. 
In some deformed nuclei,  this knowledge may be 
further extended  by E2 excitations.

Other methods to test  the nuclear surface with antiprotons detect  the 
products of $\bar{p}$  annihilation by the nucleus. The first experiment 
of this
type detected charged mesons and in this way could approximately
discriminate the captures on protons from captures on neutrons
\cite{BUG73}. On the other hand, the  recent  CERN experiments detect
residual nuclei of very low nuclear excitation \cite{LUB94}.
Radiochemistry allows to find "cold" nuclei of only one nucleon lost 
in $\bar{p}n$ and $\bar{p}p$ annihilations. In this way
one can study the ratio of $n/p$ densities. 
The surface nature of the nuclear capture processes arises from the high 
orbital quantum numbers of the annihilating $\bar{p}$ and consequently the 
radiochemical method selects the most peripheral proton or neutron orbits. 
On average there are five mesons emitted in the annihilation 
and to leave the final nucleus cold they must all avoid collision with it. 
This can be achieved only if the annihilation takes place at the extreme 
nuclear surface around a region 2.5 fm beyond the half density radius.

The advantage of the X-ray studies is that the atomic states in question 
are known. In the annihilation-product studies this is not the case,
and additional knowledge of capture states  and final-state
interactions is necessary. These two types of atomic experiments complement 
each other.

\section{THE $\bar{p}N$ AMPLITUDES AT AND BELOW  THE $\bar{p}N$ THRESHOLD} 

Atomic antiprotons scatter on  surface localized nucleons in an almost quasi-free way. 
However, in the c.m. system of $\bar{p}N$ pairs  the energy momentum relation 
is not the free one. The energy is determined by the nucleon and 
antiproton binding $E_B$, 
while momenta are described by corresponding wave functions. 
Thus one needs to extrapolate the scattering amplitudes off the energy shell.
This may generate tremendous effects  in cases of quasi-bound states or 
resonances in the  $\bar{p}N$ system.  The on shell  scattering amplitude at low 
energies may be parameterized  in terms of scattering  lengths $ a_{0}$ and  
scattering volumes $ a_{1}$  as $ a_{0}(E)+ 3 a_{1}(E) {\bf k'}{\bf k}$. 
Here, ${\bf k}$ is the c.m. momentum related to the energy E.  
Many spin and isospin states contribute to the scattering and the detailed 
structure is uncertain and model dependent. Hence, for the few-body and nuclear 
physics of antiprotons it is convenient to parameterize the data  in terms of  
an averaged effective complex length $a$  defined as 

\begin{equation}
\label{I1} 
a = <<  a_{0}(-E_B -E_{rec})+ 3 a_{1}(-E_B-E_{rec}) {\bf \nabla }{\bf \nabla }>>. 
\end{equation}
The  average is to be performed  over atomic and nuclear wave functions. 
Those  
generate distributions of the total and c.m  momenta of the pair. The total 
momentum determines recoil energy of the pair with respect to the 
rest of the nucleus  $ E_{rec}$. The relative momentum determines the 
strength of  P-wave interactions. The latter are  generated by the derivatives 
$ {\bf \nabla }$  over  relative $\bar{p}N$ coordinates. 
Such effective scattering lengths have been  commonly used to parameterize  the  
optical potential 
\begin{equation}
\label{I2} 
V^{opt}(R)= \frac{2\pi}{\mu_{NN}} \, a \, \rho (R). 
\end{equation}
Early atomic experiments determined   $a$ close to  $(-1.5 -i 2.5)$ fm \cite{ROB77}.
The depth of this potential well is $(-120 -i 200)$ MeV but it has to be stressed that 
formula (\ref{I2}) refers  only to the nuclear surface. 
The extrapolation to densities 
exceeding some 15$\%$ of the central density is not justified. 
 
In principle  $a$ is a function of energy  as it depends on the nucleon binding energies. 
For example in the  simplest nuclei  p, D, $^{3}$He  and $^{4}$He the nucleon separation
energies are roughly : 0,2,7 and 21 MeV respectively. In addition the recoil energies  
amount to several MeV.  
Thus, $\bar{p}$ atoms built with  these nuclei 
may  test the  $\bar{p}N$ scattering 
amplitudes below the $\bar{p}N$ threshold. Typical $(- E_B -E_{rec})$ energies involved 
in equation (\ref{I1}) are  0, -9,-15 and  -30 MeV.  
This region  covers most of  the energy range required in heavy  $\bar{p}$ atoms. 

Now we attempt an extraction of the length parameters  $a$  from the recent atomic 
\cite{GOT99}, \cite{AUG99},\cite{SCH92} and scattering \cite{ZAN99} experiments.
The next task is to disentangle relation (\ref{I1}) to obtain  more fundamental 
scattering parameters  $ a_{0},a_{1}$ at and below the threshold. The results are 
collected in table 1.

\begin{table}
\caption{The $\bar{p}N$ scattering parameters extracted from few body systems, 
 $a, a_{0}$ in [fm],  $ a_{1}$ in  [fm$^{3}$]  
Sources refer to atomic or scattering experiments (A,S) and calculations (C).}

\begin{tabular}{lcccc}

System          & $  a     $     &   $ a_{0}  $  &  $ a_{1} $            & $ source $  \\ \hline

$ p \bar{p}$         & 0.83 - i0.69   &  0.83(1) - i0.69(3)  & -0.4(8)-i0.64(4)     & A \cite{GOT99}          \\

$N \bar{p}$         & --  &   0.28-i0.59     &   0.02-i0.57          & C \cite{PAR94}                        
 \\

$^{2}$D $\bar{p}$    & --    & 0.22-i0.45    &  1.28-i1.25  & A \cite{AUG99}, S \cite{ZAN99}                \\

$^{3}$He $\bar{p}$        & 1.48-i2.81 & --  &  0.68- i1.25   & A \cite{SCH92}, C \cite{WGR93}    \\

$^{4}$He $\bar{p}$       & 0.77-i2.99   &  0.2(3)-i0.2(1)  &  0.3-i 1.0   & A \cite{SCH92}, C \cite{WGR93} \\

\end{tabular}
\label{table1}
\end{table}

In protonium,  $ a_{0},a_{1}$  are just {\sl the $\bar{p}p$ scattering 
length and volume}. These are related to atomic level shifts and widths 
by the Trueman formula 
$ [\Delta E-i\Gamma/2]_{l} = a_{l} \omega_{l}  + O(a/B) $. Coefficients 
$ \omega_{l}$ and higher order terms in the ratio (scattering length / Bohr radius) 
are known \cite{CAR92}. The same formula  holds for deuterium  and heavier atoms 
and in this sense the atomic levels are equivalent to the low energy scattering. 

The shifts and widths obtained recently  for the 1S and 2P states of $\bar{p} D$
atom \cite{GOT99}, \cite{AUG99} allow to calculate the $\bar{p} D$ scattering length 
$ A_{0}^{D}= (0.706(17)-i0.39(27))$ fm  and volume $ A_{1}^{D}= (3.15(33)-i3.18(19))$ 
fm$^{3}$. 
The absorptive parts of these are consistent  with  
similar values obtained from
the reaction cross sections \cite{ZAN99}, \cite{PRO00}.

To extract the $\bar{p} N$ parameters from deuterium one needs to solve the three 
body dynamics. Here, the multiple scattering series summation method of Refs.\cite{WGN85} 
is used for this purpose. For the deuteron, this method has been shown to be
very reliable. Its precursor is the Brueckner formula 
\begin{equation}
\label{I3} 
A_{0}^{D}= \frac{\mu_{ND}}{\mu_{NN}} \frac{ a_0 }{1+ a_0  \frac{1}{R_o} },
\end{equation}
which  in the static nucleon limit is obtained from  boundary
conditions  set upon a wave scattered by two centers separated by a 
distance $ R_{o}$. More careful calculations are needed, and 
then $ \frac{1}{R_{o}}$  becomes  an effective 3-body propagator. It may be  expressed in 
terms of rapidly converging partial sums of  the scattering series \cite{WGN85}. 

The values of  $ a_{0}$ and $a_{1}$ extracted from antiprotonic deuterium 
and shown in table 1  
differ strongly from the threshold values obtained from hydrogen. However, 
the former  pertain to an average $\bar{p}N$ while the latter pertain to
the $\bar{p}p$ system. To find the average $\bar{p}N$ lengths at  
threshold, the missing $\bar{p}n$ amplitudes are  calculated. The results 
obtained with the Paris potential are  shown in 
the second line of table 1. This potential has recently been tested 
against the $\bar{p}p$ and $\bar{n}p$ interactions  \cite{PAR94}. 
An independent estimate is possible for the S-wave. It follows from 
the hydrogen 1S width and Im$a_{0}(\bar{n}p)=-.83(7)$fm  extracted 
from measurements of the $\bar{n}p$ reaction cross section \cite{MUT88}. 
In this way Im$a_{0}(\bar{p}N)=-.76(5)$fm is obtained  and this number 
is  roughly 
consistent with the Paris potential calculations.   
Hence, the S-wave scattering length obtained at the threshold
and  the length  obtained from the deuteron  indicate fairly regular  
energy dependence.  On the other hand, 
a dramatic change  of the P-wave scattering volume seems to happen  
in the region between the threshold and -10 MeV.

An extension of $ a_{l}$ to lower energies may be performed  with the use  
of 2P and 3D atomic level shifts and widths in $^{3}$He, $^{4}$He, 
\cite{SCH92}. 
Again, the calculations presented in table 1  involve  the partial  
summation of multiple scattering series. This procedure is rather 
reliable for the P and D  states, \cite{WGN85}. 
To obtain $a_{0}$ we use the   $\bar{p}^{4}He$ scattering length 
extracted from  $\bar{p}^{4}He$ absorptive cross sections  
\cite{ZAN99},\cite{PRO00}.  The $ a_{0}$  given in table 1 
serves essentially as a plausible indication of the real value.  
It has been calculated with $ A_{0}^{He}= (1(1)-i0.4(.4))$fm and 
this figure doubles the error limits given in  ref.\cite{PRO00}.  
The result for  $ a_{0}$ indicates a  regular behavior  of 
the S-wave amplitude in the subthreshold energy region.
Also regular is the result for Im $a_{1}$ 
in the -15~MeV to -30~MeV region. On the other hand, Re $a_{1}$ tends 
to fall down. These conclusions are  slightly affected by 
the uncertainty of  $ a_{0}$ (indicated in the table) that induces a  
20$\%$ uncertainty in the helium values of $a_{1}$ 
(not indicated in the table).

Atomic data in light nuclei do not allow to pinpoint the partial wave 
which is 
responsible for the dramatic effect in the P wave close to threshold. 
Models for $\bar{N}N$ interactions generate fairly narrow  P wave 
resonances just above the threshold and broad quasi-bound states deep 
below the threshold.  
In particular the $^{13}P_{0}$ and $^{33}P_{0}$ waves are found to resonate
in the Paris model, \cite{PAR94}.  
We return to this question as similar effects are observed also in heavy 
$\bar{p}$ atoms.

\section{HEAVY ANTIPROTONIC ATOMS} 

In this section we discuss some  geometric properties of the antiproton 
nuclear scattering and capture. These are: 

$i)$. A saturation of level widths $\Gamma(Z)$ as the atomic number increases. 

$ii)$. A scaling of the ratio $\Delta E / \Gamma$(R) as the nuclear radius increases. 

$iii)$. An effect of $\bar{p}N $ force range that increases with the atomic angular 
momentum.  

With some control over these simple properties one  finds anomalies and attempts  
the construction of  phenomenological optical potentials.  With the potentials one 
can compare the results of the X-ray measurements with the data on $\bar{p}$ single 
nucleon  capture. Consistency of these experiments would give a signal that antiprotons 
make  a valid tool to study properties of the distant nuclear surface. 
Finally the nuclear potentials may offer a check for the underlying 
$\bar{p}N $ scattering amplitudes. 

\subsection{Geometric properties of absorptive interactions} 

The three related, but different geometric effects are now discussed on the basis of the 
extended atomic data. 

$i)$.  The saturation of the scattering length has been recently discovered in light 
nuclei with the relation of scattering lengths   Im $A(\bar{p}He) < $ Im $A(\bar{p}p)$, 
\cite{ZAN99},\cite{PRO00}.  
A similar effect is seen  with the atomic level widths, on 
a broader Z scale. 
To obtain  an equivalent scattering parameter let us divide  the experimental 
widths by  normalization factors for the  atomic densities. 
This prescription  normalizes the widths to the same number of antiprotons 
approaching the nuclear surface.  The results  for n=6, l=5 level widths 
are shown in Fig.1.  Most of these  atoms  have  several isotopic states.  
On average one detects an initial increase and  a fall at 
the heaviest  Te atom. The width for this last state has been obtained indirectly 
by the E2 mixing effect. This state is of independent interest as it reflects the 
largest atomic-nuclear overlap ever tested and 
may be interpreted as a Coulomb assisted nuclear $\bar{p}$ state. 
For each element the scaled widths display isotopic differences and these  reflect 
a change of the nuclear size. The  differences in the  Bohr radii 
of the atomic states have  already been accounted for.

\begin{figure}[!hbt]
\includegraphics[width=0.9\textwidth]{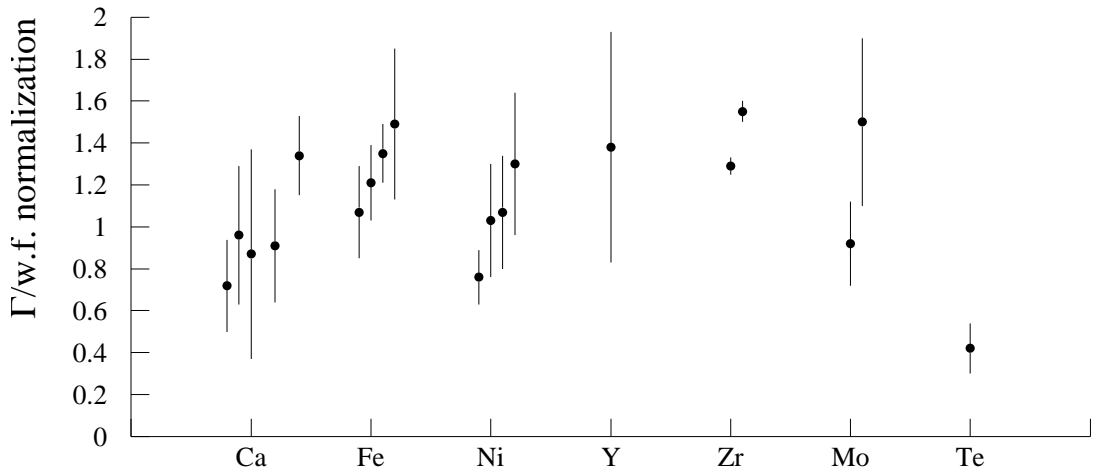}
\caption{Experimental widths of  n = 6, l =5 atomic levels, scaled by the 
normalization factors of the corresponding  wave functions. Arbitrary units. 
The data: Ca,Fe,Ni,Zr,Te from \cite{TRZ00},Fe,Y,Zr from \cite{ROB77} 
(Roberson), Mo 
from \cite{KAN86}.}
\label{fig1}
\end{figure}

The saturation indicated above is due to strong damping of the initial 
$\bar{p}$ wave in subsequent $\bar{p}N$ collisions. It may be explained in 
simple terms of the two center formula  (\ref{I3}),  if Im$ a_{0}$ 
is allowed to increase and $ R_{o}$ stays  small enough.

$ii)$.  Another scaling effect may be seen in the ratio $\Delta E / \Gamma (Z)$.
This scaling is more subtle than the saturation of  widths. 
The lower level shifts are predominantly repulsive  despite the fact that 
the optical potential may be an attractive one. The repulsion is related to strong 
damping of the atomic  wave function as  $\bar{p}$ penetrates the nuclear interior. 
Such damping 
results in a large gradient of the wave function and  pushes up the  
kinetic energy i.e. generates an effective repulsion. To see the effect we 
divide level shifts by level widths. The latter set the scale of the atomic-nucleus  
overlap. On this scale the shift becomes smaller as the size of the nucleus increases 
and the gradients become weaker. The result is plotted in Fig.2.

\begin{figure}[!hbt]
\begin{center}
\includegraphics[width=9cm]{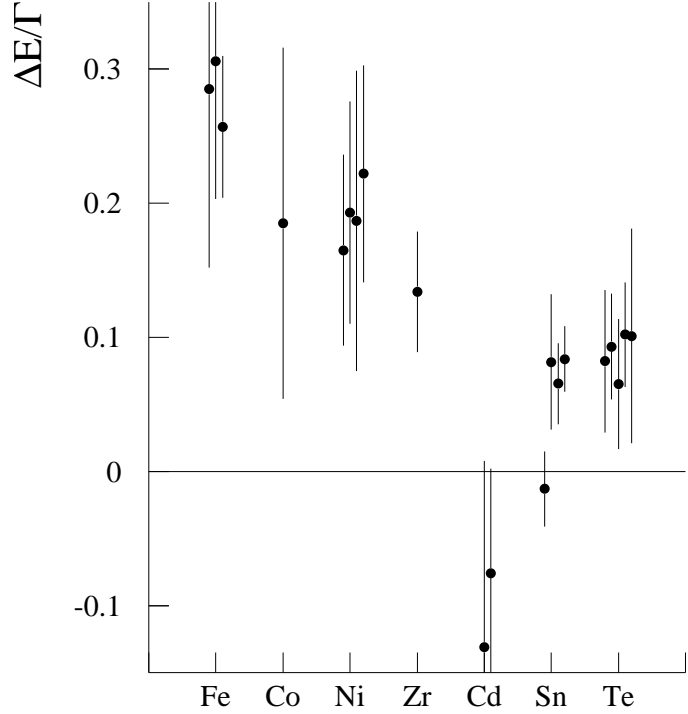}
\end{center}
\caption{The ratio of experimental $\Delta E / \Gamma $.
The data: Fe,Ni,Zr,Cd,Sn,Te from \cite{TRZ00}, Fe,Y,Zr from \cite{ROB77} 
(Roberson).}
\label{fig2}
\end{figure}

This  scaling of $\Delta E / \Gamma $ may be  also reproduced in simple terms 
of formula (\ref{I3}), if the length $a_{0}$ is dominated by the absorptive part 
and the size of the system $R_{0}$ is allowed to increase. Fig.2  indicates an 
average behavior of this ratio.  
On top of it  some shifts  do not follow the trend and become negative. 
Two effects may contribute to this anomaly: in Cd the overlap (and thus the absorption)
is small while in $^{112}$Sn and $^{106}$Cd one has loosely bound 
valence protons.

$iii)$.  An important  geometric effect is related to the  $\bar{p}N $ force range.  
It enters the potential profile in eq.(\ref{I2}) which is usually  presented in a 
folded density form  
\begin{equation}
\label{H1}
\rho (R) = \int \rho_{o}({\bf R-r }) \upsilon({\bf r}) d{\bf r}
\end{equation}
where $ \rho_{o} $  is a " bare" nucleon density  and  $\upsilon({\bf r})$ is a 
formfactor describing the range in $\bar{p}N$ interactions. 
The impact of the force range is determined, essentially, by the m.s. radius  of 
the formfactor  $r_{o}= \sqrt{< r^{2}>}$. 
To see the effect let us calculate the atomic level width 
with the formula  
\begin{equation}
\label{H3}
  \Gamma_{s}/2 ={\rm Im} 
       \int\mid\Psi_{\bar{p}}( R)\mid^{2}
       V^{opt}(R) d{\bf R} \quad  \approx const < R^{2l}>.
\end{equation}
In a state of angular momentum  $l$ the wave function at short distances 
$\Psi_{\bar{p}}(R) \sim  R^{l}$, hence  the width is  roughly 
proportional to the $ 2l$-th moment of the folded  density 
$< R^{2l}>$. The latter may be 
related to the bare nucleon density moment $< R_{o}^{2l}>$. To the leading 
order in $< r^{2}>$ this relation is  
\begin{equation}
\label{H4}
  < R^{2l}> =  < R_{o}^{2l}>  + < R_{o}^{2l-2}>< r^{2}> \lambda_{l} + .....
\end{equation} 
Coefficients 
$ \lambda_{l} $ rise  quickly with l. For l=l,2,.. one has  $ \lambda_{l} = 
1 , 10/3, 7 , 12 , 55/3 ..$  and the second term in 
eq. (\ref{H4}) increases rapidly.
The effect of force range becomes more and more important, particularly in the 
high l states which may be created in heavy atoms.

\subsection{The optical potential} 
 
Phenomenological  potentials that are linear in the nuclear density and 
the scattering matrix are characterized  by at least three parameters: 
the complex  length  $a$  and an interaction range  $r_o$. 
Early best-fit  potentials assumed  $r_{o}$ equal to the proton charge 
radius $r_{ch}$  which is roughly  the $\bar{N}N$ annihilation radius. 
The values $a \approx (-1.5 -i 2.5)$ fm were obtained \cite{ROB77}. 
A recent choice of ref.\cite{BAT97} is the extreme zero range limit. 
Now this way   $a= (-2.5(2) -i 3.5(3))$ fm becomes larger to  compensate  
for the shorter force range.  These values of   $a$ have been determined 
mostly by the light atoms  where the data are more precise. Attempts to 
extract separate proton and neutron values yield  uncertain results 
\cite{BAT97}. The new data from PS209 \cite{TRZ00} offer a chance and 
a challenge to describe interactions  
in the whole periodic table, in particular for the neutron rich nuclei.

The old problem in the optical potential studies is that the sign of the 
effective Re $a$ differs  
from the signs of  averaged scattering lengths and volumes. 
The best-fit Re $a$ are  attractive  while  Re $a_{0}$ and  Re $a_{1}$ at 
the threshold are repulsive. It is seen in table 1 that below the 
threshold these quantities are also repulsive. The 
attraction observed in $\bar{p}$ atoms and in the low energy scattering 
\cite{GAR86} is due to collective nuclear effects and 
a more subtle description of the $\bar{p}N$ scattering amplitudes. The basis 
to describe these effects exists in the standard approach which  generates 
the optical potential in terms of the half-off shell scattering matrix  
$t(r)$.  
Instead of equation (\ref{I2}) which uses the effective scattering lengths 
$a$,   one uses  $t(r)$  and folds  it over the nucleon density  
\begin{equation}
\label{H5}
V^{opt}(R)= \frac{2\pi}{\mu_{NN}} \int \rho_{o}({\bf R-r }) t({\bf r}) d{\bf r}.
\end{equation}  
Since $a$ and $t(r)$ are related  by $a = \int t(r) d{\bf r}$,   
formula (\ref{I2}) can  approximate formula (\ref{H5}) only in a 
simple case of regular, monotonic  $t(r)$.  
This does not happen. At large distances  $ t(r) $
is given by the $\bar{p}N$ potential and  the  average potential  in this 
system is attractive.  On the other hand, at smaller  distances 
Re $t(r)$ changes sign and becomes repulsive. 
This sign change is due partly to the annihilation and partly to 
$\bar{p}N$  quasibound states. Such a mechanism generates the 
repulsive on average scattering lengths and volumes.  
The net result of the folding  (\ref{H5}) is that 
at large distances  Re $V^{opt}(R)$ becomes attractive while at 
short distances it is repulsive. 
Several model calculations reproduce such an effect. However, the detailed 
calculations are uncertain due to  strong cancellations involved in the 
outlined procedure. The problem is aggravated by  $\bar{N}N$ model 
uncertainties and   technical questions: uncertainties in the full 
off-shell versus half-off shell extension, difficult description of 
the nuclear surface region  and an early onset of nuclear many body 
effects. Several involved calculations were undertaken in the 
former decade and  all indicate uncertainties due to such effects 
\cite{POT80}.
One certain conclusion is that Re $V^{opt}$ is a very complicated, 
possibly  energy dependent and  non-local structure. In some states 
it is attractive while in other states  is may be repulsive.  

Below a new best-fit potential is presented. It stems from the early 
best-fit potentials. The basic changes are:

$(1)$.  Light nuclei are described by two parameter Fermi densities and not 
by the harmonic oscillator  densities which  generate incorrect asymptotics. 
For other nuclei either the 
electron scattering or  muonic atom data are used, whichever yield better 
fit. For neutrons the $(r_{n}-r_{p})$  differences are taken from other 
experiments or interpolated  \cite{BAT89}  and implemented into a change 
of  the  diffuseness parameter. 
 
$(2)$.  The starting range parameter was  $r_{ch}$. Next, this condition was relaxed. 

$(3)$.  A constant length $a $ was assumed. Next,  some dependence on the 
          separation energies indicated by  the lightest antiprotonic 
          atoms was allowed.

$(4) $.  The data base is extended by new results \cite{TRZ00}. Some
    150  lower shifts, lower widths and upper widths are used in the carbon 
    till uranium  region  of the periodic table.
 
The best choice for a constant  $ a=( -1.10(5) -i1.85(5)) fm$  is suggested 
by light atoms. The fit was obtained with  $ \chi ^2/F $ = 1.19 ( for $ Z< 17$), 
1.46( for $ Z<38$ ), 2.72 ( for $Z<53$ ) and 3.20 (for all Z ). This fit is 
excellent for the old data and light atoms but the representation of heavy 
atoms is poor. The improvements were looked for in the energy dependence 
of $a$  indicated by results of table 1. 
In particular  an enhancement of absorption on loosely bound nucleons 
was expected. 
For the data in existence the separation energies of the valence nucleons 
span the  region from  3.6 MeV  to 18.5 MeV.  Within this region 
only minute  changes in the neutron Im$a_{n}$ are allowed by the data. 
On the other hand a  100 $\%$ increase of Im$a_{p}$ on weakly  bound 
protons ( with separation energies of less than 8 MeV ) is allowed  and 
favoured by the data. The best fit  
$ \chi ^2/F $ = 2.5 ( for all Z) is obtained in this way. This shows  
that the resonance effect indicated by the 
light atoms is likely to be attributed to the $\bar{p}p$ system.

With the antiprotonic atoms the interesting region of subthreshold energies  
of less than 10 MeV cannot  easily be reached. The antiproton annihilations 
involve not only the  valence but  also other nuclear shells.  On the 
other hand, the single nucleon capture processes,  localized at the extreme 
surface, happen mostly on the valence nucleons. As discussed in next section 
the energy dependence seen there seems to be more  drastic.

What is found with the recent data \cite{TRZ00} is that for light nuclei 
a best fit $a$ may be obtained for a number of  ranges  $r_{o}$, however, 
a fit over all the periodic table favours  $r_{o}$ slightly larger than 
$r_{ch}$. 
The best fit may be obtained with $r_{o}$ = 1 fm but the $ \chi ^2/F $ is 
changed only marginally to  2.4.

\subsection{Relation of the X-ray and the  single nucleon capture experiments} 

The atomic levels test nuclear densities in surface layers of 
some 2.5 fm in  depths. The mean radii correspond roughly to  
R+1.2 fm (lower levels widths), R+1.5 fm (upper level widths) and 
R+2.5 fm (single nucleon captures), where R is a half density radius of 
the nucleus. The motivation for the radiochemical capture experiments was to  
study the relative $\bar{p}n  / \bar{p} p $  capture rates  $\sigma_{n/p}$ 
at far nuclear peripheries. From these rates the ratios of neutron to proton 
densities were extracted. Results  indicated neutron halos in most 
of the studied medium and heavy nuclei \cite{LUB94}. 
The $\bar{p}n  / \bar{p} p $ capture ratios may be also obtained from 
the atomic level widths. These are defined as 
$\sigma_{n/p}= \Gamma^{exp}/ \Gamma^{prot}-1 $ where $\Gamma^{prot}$ 
is a  partial level width that corresponds  to the annihilation on a proton. 
This width has to be calculated with the best-fit potential based on known 
(in principle)  proton density. Few results, and comparison of the two 
experiments are shown in table 2. 

\begin{table}
\caption{The relative capture rates $\sigma_{n/p}$ obtained from the lower and upper level 
widths \cite{TRZ00}, $a=(-1.1-i1.85)$fm  is used. The last column shows $\sigma_{n/p}$  
obtained via the  single nucleon captures  \cite{LUB94}.}
\begin{tabular}{lccc}

Atom         & lower    &  upper  &  capture        \\ \hline

$ ^{48}$Ca    &  --      &  1.58(28)  & 2.62(30)      \\
$ ^{96}$Zr    & 0.96(9)  &  1.54(29)  & 2.6(3)        \\
$ ^{116}$Cd   & 1.64(49) &  2.67(61)  & 5.00(21)      \\
$ ^{124}$Sn   & 1.80(10) &  2.46(39)  & 5.0(6)        \\
$ ^{128}$Te   & 1.03(19) &  2.68(56)  & 4.2(1)        \\   \hline \hline

$ ^{106}$Cd    &  1.65(80)  &  5.13(80)  & .5(1)       \\
$ ^{112}$Sn    &  1.91(13)  &  2.45(49)  & .79(14)     \\
\end{tabular}
\label{table2}
\end{table}

Nuclear physics  predicts the neutron/proton density ratios to increase 
at large distances. This happens predominantly as a result of the Coulomb 
barrier, subject 
to differences in  the separation energies and centrifugal  barriers. 
Such a behaviour  is borne out by the three  complementary measurements: 
the lower level width, the upper level  width and the single nucleon capture. 
These test more and more extreme surface regions. 
Consistency of the two  experiments is indicated  in the upper five lines, 
several additional cases exist in the data. The errors attributed to  
the atomic values are statistical only. In addition  there exists sizable 
uncertainty due to the calculation of $\Gamma^{prot}$. These make a smooth 
behaviour of the $\sigma_{n/p}$ ratios to be an additional, strong constraint 
on the  nuclear densities and the structure of optical potentials.  

The two lowest lines of table 2 indicate a strong disagreement between the 
two experiments. Again the interesting point is that these results 
are characterized by low proton separation energies of 7.35 MeV in 
$ ^{106}$Cd and 7.54 MeV in $ ^{112}$Sn.
In the upper sector of the table  the proton separation 
energies are close to  or larger then 10 MeV.

\section{UNSOLVED QUESTIONS} 

Several new results  suggest an interesting physics likely to happen in the 
antiproton interactions on loosely bound protons. These are : 
anomalies in the relative neutron/proton single nucleon capture ratios, 
attractive lower shifts in $ ^{106}$Cd, $ ^{112}$Sn, the  enhancement 
of $\bar{p}p$ absorption 
for  negative but close to threshold energies and the  difference 
between scattering volumes  obtained from the protonium  and from the 
deuteronium.  

All these,  indicate that a particular role in the interaction may be  
played by the $^{13}P_{0}$ wave. This state, of vacuum quantum numbers, 
is characterised by very strong  tensor forces generated by the 
pion exchange.  If the annihilation in the  spin triplet states  is weak  
these 
forces generate a fairly narrow resonance just above the threshold and 
a  quasi-bound  state well below it. To understand the 
effects  discussed here  both these states seem to be required. 
However, there exists a  difficulty in this  interpretation and in 
the related experimental search. Such a resonant state has a large 
radius of
1-2 fm. It cannot be  built in  nuclear systems. One 
has to search for it in very   low density situations. Those are 
met, for instance,  in the single nucleon $\bar{p}p$ captures or in the 
antiprotonic hydrogen.  In the latter  case,  the effect of $^{13}P_{0}$ 
state is  seen in the fine structure of the 2P states \cite{GOT99}, 
\cite{AUG99}. A large scattering volume in this state  
has been  confirmed  in this way. Unfortunately the latter is almost 
independent of the annihilation models. To  understand 
the $^{13}P_{0}$  resonance better, a few MeV step below the threshold 
is required. This may be  realised in terms of nuclear experiments. 

The relation to other  recently found, resonant-like phenomena is  not 
transparent as yet. 
The dip found in the $ e^{+}e^{-}$ annihilation just below the $\bar{p}p$ 
threshold \cite{ANT98}, might be attributed to the $^{13}S_{1}$ wave. 
This effect is at least consistent with the trend of the S wave absorption  
found in deuteronium and protonium.  Another recent finding of a destructive 
interference in the  $\bar{n}p$ scattering just above the threshold 
\cite{IAZ00}  may be attributed to an effect of  the $^{33}P_{0}$ wave. 
An extrapolation of this effect to the subthreshold region is uncertain, at 
this moment.  

To elucidate these questions,  new experiments would be useful: 
the elastic  $\bar{p}D $ scattering at few MeV energies, resolution of 
the LS splitting in upper atomic levels and studies of hadronic atoms 
built on nuclei  with very loosely bound nucleons.

Acknowledgements: The PS209 team is thanked for the encouragement and
collaboration. This work  was partly sponsored by the Deutsche 
Forschungsgeminschaft, Bonn.

\end{document}